\documentclass[journal]{IEEEtran}
\usepackage{amsmath,amsfonts}
\usepackage{algorithmic}
\usepackage{algorithm}
\usepackage{array}
\usepackage[caption=false,font=normalsize,labelfont=sf,textfont=sf]{subfig}
\usepackage{textcomp}
\usepackage{stfloats}
\usepackage{url}
\usepackage{verbatim}
\usepackage{graphicx}
\usepackage{cite}
\begin{document}

\title{A summary of the routing algorithm \\ and their optimization,performance }

\author{Xunchi Ma \\ H00392669 \\ 21012100015}

\markboth{Personal essay for the Principles and practices of Interconnection Network}%
{Shell \MakeLowercase{\textit{et al.}}}

\maketitle

\begin{abstract}
This essay provides a comprehensive analysis of the optimization and performance evaluation of various routing algorithms within the context of computer networks. Routing algorithms are critical for determining the most efficient path for data transmission between nodes in a network. The efficiency, reliability, and scalability of a network heavily rely on the choice and optimization of its routing algorithm. This paper begins with an overview of fundamental routing strategies, including shortest path, flooding, distance vector, and link state algorithms, and extends to more sophisticated techniques.
\end{abstract}


\section{Introduction}
\IEEEPARstart{T}{his} paper presents a comprehensive summary of routing algorithms, including their optimization and performance metrics. With the expansion of network scale, traditional point-to-point (P2P) communication has become impractical. This limitation is not confined to computer networks, which are widely recognized, but extends to Network on Chip (NoC), optical networks, social networks, and other relevant domains. To improve network performance metrics such as communication bandwidth, latency, delay, and packet loss ratio, it is imperative to analyze the network structure, topology, and routing algorithms.

This study consists of some typical routing algorithm. Most of them are the examples in the course we have learnt before. All of these routing algorithm are not perfect. They have their own advantages and drawbacks, which decide the specific application scenarios. despite their simple prinple, they have formed the base of the modern routing algorithm. 

This study also encompasses an exploration of several advanced routing algorithms. While most of these algorithms are currently in the research phase and have not been deployed in production environments, they offer valuable insights into the evolving trajectory of routing algorithm development. This investigation aims to contribute significantly to the field by forecasting future trends and fostering advancements across the industry.

\section{Some typical routing algorithm \\ and their pros \& cons}
\subsection{Shortest Path Algorithm}
The Shortest Path Algorithm is a fundamental concept in the field of computer science and networking, aimed at finding the shortest path between two points in a graph. This graph can represent various physical, social, or abstract networks, where nodes represent entities and edges represent the connection or distance between these entities. The shortest path is the one that minimizes the sum of the weights of the edges traversed between the source and destination nodes. There are several algorithms designed to solve the shortest path problem, with Dijkstra's algorithm and the Bellman-Ford algorithm being among the most prominent.
\paragraph{Dijkstra's Algorithm}
 Proposed by Edsger W. Dijkstra in 1956, this algorithm solves the shortest path problem for a graph with non-negative edge weights, producing a shortest path tree. This tree contains the shortest path from the source node to all other nodes in the graph.
 
How it Works: Dijkstra's algorithm maintains a set of nodes whose shortest distance from the source is already known and a set of nodes whose shortest distance is not yet determined. Initially, the distance to the source is zero and infinity for all other nodes. The algorithm repeatedly selects the node with the smallest known distance from the source, updates the distances to its neighbors, and repeats the process until all nodes have been processed.\\\\
\textbf{Pros:}
\begin{itemize}
	\item It guarantees to find the shortest path in graphs with non-negative edge weights.
	\item It is relatively straightforward to implement.
\end{itemize}
\textbf{Cons:}
\begin{itemize}
	\item It cannot handle graphs with negative edge weights.
	\item Its computational complexity can be high for dense graphs without optimizations like priority queues.\\
\end{itemize}
\begin{figure}[!t]
	\centering
	\includegraphics[width=2.5in]{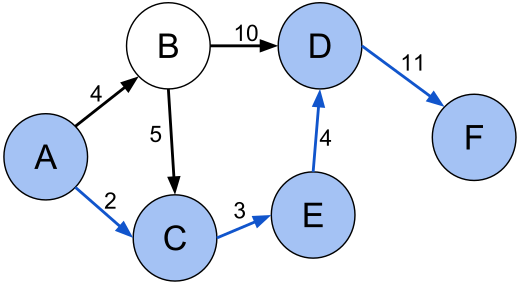}
	\caption{Shortest path with direct weights.\cite{shortest}}
	\label{Shortest_path_with_direct_weights}
\end{figure}
\paragraph{Bellman-Ford Algorithm}
The Bellman-Ford algorithm extends the capability to handle graphs with negative edge weights, allowing it to detect negative weight cycles in the graph.

How it Works: It relaxes the edges of the graph repeatedly, updating the cost to reach each node from the source if a cheaper path is found. This process is repeated for each edge, and the algorithm checks for negative cycles, which occur if a cost can be further reduced after sufficient iterations.\\\\
\textbf{Pros:}
\begin{itemize}
	\item Capable of handling graphs with negative edge weights.
	\item Can identify negative cycles in the graph.
\end{itemize}
\textbf{Cons:}
\begin{itemize}
	\item It is slower than Dijkstra's algorithm due to higher time complexity.\\
\end{itemize}
Shortest Path Algorithms are widely applicable in:
\begin{itemize}
	\item \textbf{Network Routing:} Optimizing data packet paths in computer networks.
	\item \textbf{Geographic Mapping:} Determining shortest routes in mapping and navigation systems.
	\item \textbf{Resource Allocation:} Efficient route planning in logistics and supply chain management.
	\item \textbf{Social Networking:} Analyzing connections in social networks.
\end{itemize}
\subsection{Randomly Pick Direction}
The Randomly Pick Direction (RPD) routing algorithm represents an unconventional approach in the landscape of routing strategies, particularly within the realms of ad hoc networks, mesh networks, and certain types of distributed systems. Unlike deterministic or optimized routing algorithms that seek the most efficient path based on specific metrics (like shortest path, least congestion, or highest throughput), RPD bases its routing decisions on randomness. Here, we delve into the mechanics, advantages, and potential drawbacks of the RPD routing algorithm.

The core principle of the RPD routing algorithm is its reliance on random selection to determine the next hop for a packet. When a packet arrives at a node, instead of calculating the best path to the destination, the node randomly selects one of its available neighbors as the next hop. This process repeats at each intermediate node until the packet reaches its intended destination or a predefined maximum number of hops is exceeded.\\\\
\textbf{Pros:}
\begin{itemize}
	\item Simplicity: The algorithm is straightforward to implement as it does not require complex calculations or maintenance of detailed routing tables. This simplicity is particularly beneficial in rapidly changing or unpredictable network environments.
	\item Load Balancing: By randomly distributing the packets across various paths, RPD can naturally balance the load among different nodes and links, potentially reducing congestion on popular routes.
	\item Robustness: The randomness inherent in RPD provides a degree of resilience against certain types of network failures or attacks, as there is no fixed path for malicious entities to target or for failures to disrupt.
\end{itemize}
\textbf{Cons:}
\begin{itemize}
	\item Inefficiency: Random path selection is inherently less efficient than algorithms that optimize for specific metrics. This can lead to longer paths, increased latency, and higher packet loss rates, especially in dense or complex networks.
	\item Unpredictability: The performance of RPD can be highly variable and unpredictable, making it difficult to guarantee service quality or meet specific performance criteria.
	\item Wastefulness: The algorithm may generate excessive network traffic due to the non-optimal paths taken by packets, leading to wasted bandwidth and energy, particularly in wireless networks.\\
\end{itemize}
Applications
\begin{itemize}
	\item Dynamic and Ad Hoc Networks: In environments where network topology changes frequently, the simplicity and adaptability of RPD make it a suitable choice.
	\item Load Testing and Simulation: RPD can be used to simulate worst-case scenarios or to test the robustness of networks against random traffic patterns.
	\item Initial Route Discovery: In some hybrid routing schemes, RPD can serve as an initial route discovery mechanism before more optimized paths are established.\\
\end{itemize}
The Randomly Pick Direction routing algorithm, with its unique approach of utilizing randomness in routing decisions, offers a trade-off between simplicity and performance. While not suited for all applications, particularly those requiring high efficiency and predictability, RPD finds its niche in scenarios where network flexibility, robustness, and simplicity are prioritized over optimal performance. Its role in the broader ecosystem of routing algorithms highlights the diversity of strategies available for managing data flow across complex networks.
\subsection{Dimention-Ordered Routing}
Dimension-Ordered Routing (DOR) is a deterministic routing strategy commonly employed in mesh and hypercube network topologies, particularly within the context of parallel computing and network-on-chip (NoC) architectures. This algorithm is characterized by its methodical approach to routing, where packets are forwarded through network dimensions in a specific order until they reach their destination. The most common forms of Dimension-Ordered Routing include XY routing (for 2D meshes) and XYZ routing (for 3D meshes), though the concept extends to higher dimensions in more complex topologies.

Principle: In DOR, each node in the network is assigned a unique coordinate based on its position within the network's dimensional grid. Routing decisions are made based on the coordinate difference between the source and destination nodes, with packets being forwarded along one dimension at a time until the coordinate in that dimension matches the destination's.
\begin{figure}[!t]
	\centering
	\includegraphics[width=2.5in]{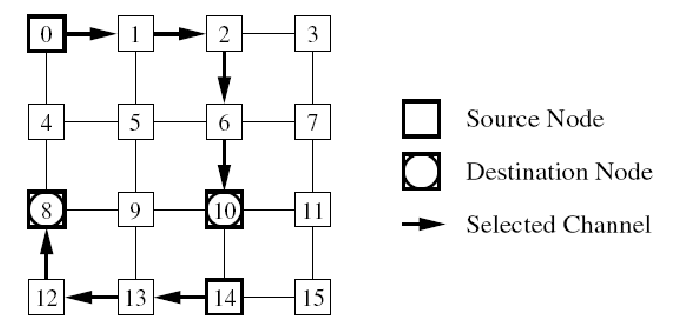}
	\caption{Routing example for dimension order routing on a 2D meshs.\cite{dimension}}
	\label{Routing-example-for-dimension-order-routing-on-a-2-D-mesh}
\end{figure}
Operation:
\begin{itemize}
	\item XY Routing (2D Mesh): Packets first move along the X dimension until they align with the destination's X coordinate, and then proceed along the Y dimension.
	\item XYZ Routing (3D Mesh): Similar to XY routing, but packets also move along the Z dimension if necessary, following the X and Y movements.
\end{itemize}

\textbf{Pros:}
\begin{itemize}
\item Simplicity: The algorithm is straightforward, making it easy to implement and understand. It requires minimal computational overhead, contributing to lower latency in routing decisions.

\item Determinism: Since the path is determined solely by the source and destination coordinates, the routing behavior is predictable, which simplifies deadlock and congestion management.

\item Scalability: DOR scales well with network size, as the routing logic does not fundamentally change with the addition of more nodes or dimensions.
\end{itemize}
\textbf{Cons:}
\begin{itemize}
\item Path Rigidity: The deterministic nature of DOR means that the paths are fixed, which can lead to congestion if certain routes become heavily trafficked.

\item Suboptimal for Non-Uniform Traffic: In cases where network traffic is not uniformly distributed, DOR may not utilize all available paths efficiently, leading to underutilization of network resources.

\item Vulnerability to Failures: The fixed routing paths mean that any failure in the network can significantly impact communication, requiring additional mechanisms for fault tolerance.
\end{itemize}
Application:
\begin{itemize}
	\item Parallel Computing Systems: Where predictable and efficient communication patterns are essential for performance.
	\item Network-on-Chip (NoC) Architectures: In multicore processors, where a predictable and low-latency communication scheme is crucial for inter-core communication.
	\item Embedded Systems and IoT Devices: Where computational simplicity and predictability are more critical than adapting to varying traffic patterns.
 \end{itemize}
\subsection{Oblivious Routing}
Oblivious Routing is a network routing strategy characterized by its use of predetermined paths for packet transmission, regardless of the current state or traffic conditions within the network. This approach contrasts with adaptive routing strategies that dynamically change paths based on network congestion or failures. Oblivious routing algorithms determine the path between a source and a destination solely based on the identities of these endpoints and do not alter the routing decisions in response to network traffic changes.

Oblivious routing involves directing traffic from its origin to its destination along preset paths, with the routing decisions made without regard to the current traffic levels. Although determining the best oblivious routing strategy is generally unfeasible for all network layouts, our findings demonstrate that it is feasible for the structured layouts commonly found in data center networks.\cite{9796682}\\
Principles of Oblivious Routing:
\begin{itemize}\item Path Determination: Routes between any pair of nodes are predefined and do not change in response to network conditions. These paths can be computed offline, based on the network topology and possibly aiming to optimize certain performance metrics averaged over a range of traffic patterns.
	\item Load Distribution: A key goal of oblivious routing is to distribute traffic evenly across the network to prevent congestion on any single path. This is often achieved through the use of randomized or round-robin selection among multiple precomputed paths for each source-destination pair.
\end{itemize}
\textbf{Pros:}
\begin{itemize}
\item Simplicity: The oblivious nature of the routing simplifies the implementation, as the path computation is done offline and does not require complex, real-time decision-making algorithms.
\item Predictability: Since the routes do not change in response to traffic, network behavior under given traffic patterns is predictable, facilitating easier network analysis and planning.
\item Scalability: Oblivious routing can scale well with network size because it avoids the overhead of dynamic route computation and state information exchange.
\end{itemize}
\textbf{Cons:}
\begin{itemize}
	\item Inefficiency under Specific Conditions: Oblivious routing may not handle sudden, localized spikes in traffic optimally, as it cannot reroute traffic to less congested paths in real-time.
	\item Suboptimal for Varying Traffic Patterns: Since paths do not adapt to changing network conditions, oblivious routing may not be as efficient as adaptive routing in environments where traffic patterns vary widely and unpredictably.
	\item Potential for Congestion: If the predetermined paths are not well-chosen, or if traffic patterns change significantly from those anticipated during path computation, parts of the network may become congested.
\end{itemize}
\textbf{Application:}
\begin{itemize}
\item Data Center Networks: In environments with well-understood and relatively stable traffic patterns, oblivious routing can provide a good balance between simplicity and performance.
\item Parallel Computing: Oblivious routing can be effective in parallel computing environments where communication patterns are predictable and can be optimized offline.
\item Networks with Predictable Traffic: Any network scenario where traffic patterns are known in advance and do not exhibit significant fluctuations may benefit from the predictability and simplicity of oblivious routing.
\end{itemize}
Oblivious routing offers a trade-off between the simplicity and predictability of fixed routing paths and the potential inefficiency in the face of dynamic, unpredictable traffic patterns. Its suitability varies depending on the specific requirements and characteristics of the network, including traffic stability, the need for real-time adaptability, and the complexity of managing dynamic routing protocols. In contexts where traffic patterns are well-understood and relatively stable, oblivious routing can significantly simplify network design and operation while providing adequate performance.
\subsection{Flooding}
Flooding is a fundamental routing technique used in computer networks to disseminate information from one node to all other nodes in the network without the need for a predefined route. This brute-force method ensures that a packet sent from a source node is delivered to all possible destinations by transmitting it to every neighbor except the one it was received from. Flooding is simple and does not require complex route discovery or maintenance protocols, making it useful in certain network scenarios.\\
How Flooding Works:
\begin{itemize}
\item Packet Transmission: When a node receives a packet intended for flooding, it duplicates the packet and sends it to all its directly connected neighbors, except the node from which it received the packet.
\item Termination Condition: To prevent packets from circulating indefinitely, flooding typically employs one of two mechanisms:\\
a) A hop count limit (TTL - Time To Live), which decrements at each node. The packet is discarded when the TTL reaches zero.\\
b) Sequence numbers to identify and discard duplicates, preventing a node from forwarding the same packet more than once.
\end{itemize}
\textbf{Pros:}
\begin{itemize}
\item Simplicity: Flooding does not require sophisticated routing algorithms or knowledge of the network topology, making it easy to implement.
\item Robustness: It ensures message delivery as long as there is at least one path between the source and destination, making it highly reliable in highly dynamic or disrupted network environments.
\item Low Latency: Since packets take all possible paths to the destination, they are likely to find the shortest path, potentially reducing latency for critical applications.
\end{itemize}
\textbf{Cons:}
\begin{itemize}
\item High Network Traffic: Flooding generates a large amount of redundant network traffic, which can lead to congestion and deplete network resources.
\item Inefficiency: Sending the same packet through multiple paths without considering the network's state can be highly inefficient, especially in large or dense networks.
\item Security Concerns: The indiscriminate broadcasting nature of flooding can pose security risks, making it easier for malicious entities to intercept or inject packets.
\end{itemize}
\textbf{Application:}
\begin{itemize}
\item Route Discovery: In some network protocols, such as Dynamic Source Routing (DSR) in ad hoc networks, flooding is used for route discovery phases before switching to more efficient routing for data transmission.
\item Broadcasting: Flooding is an effective method for distributing information to all nodes in the network, useful in applications requiring broad dissemination, such as updates or emergency alerts.
\item Distributed Systems: In certain distributed algorithms, flooding is used to ensure that all nodes receive critical control messages or synchronization signals.
\end{itemize}
Flooding, with its simplicity and robustness, is an essential routing technique in specific network scenarios where reliability and broad reach are more critical than efficiency and resource consumption. Despite its drawbacks, such as high network traffic and potential security vulnerabilities, it remains a valuable tool in the networking toolkit, especially for applications requiring universal message dissemination or in networks where the topology changes frequently.
\begin{figure}[!t]
	\centering
	\includegraphics[width=2.5in]{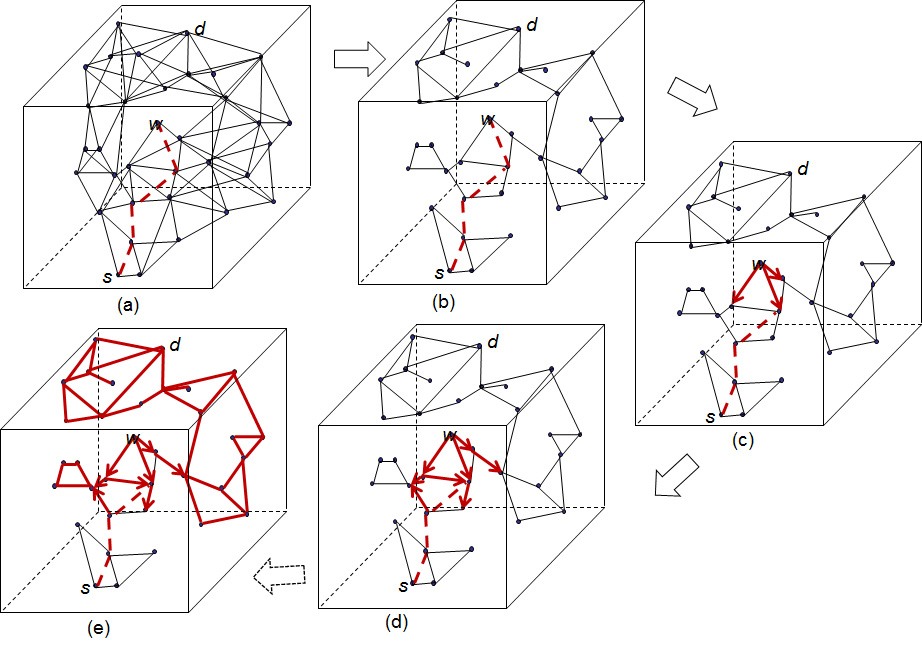}
	\caption{how floofing works.\cite{ABDALLAH2016264}}
	\label{flooding}
\end{figure}

\subsection{Distance Vector Routing}
Distance Vector Routing is a classical routing algorithm used in packet-switched networks, where routers determine the best path to each destination based on the distance to those destinations. The "distance" is typically measured in terms of hops, but it can also account for other metrics like latency, bandwidth, or cost. This algorithm is based on the principle of dynamic programming and is known for its simplicity and ease of implementation.

How Distance Vector Routing Works:
\begin{itemize}\item Distance Vectors: Each router maintains a table (distance vector) that contains the best known distance to every destination in the network and the next hop router on the best path to that destination.
	\item Periodic Updates: Routers periodically exchange their distance vectors with their immediate neighbors. Each router updates its own distance vector based on the information received, calculating potential new paths that might be better than the current ones. The update is based on the Bellman-Ford algorithm, which helps in finding the shortest path.
	\item Metric Calculation: The cost to reach a destination is calculated using the metric of distance. When a router receives a distance vector from a neighbor, it adds the cost to reach that neighbor to the costs in the distance vector. If this sum for a destination is lower than the known cost, the router updates its distance vector with this new lower cost and the corresponding next hop.
 \end{itemize}
 \textbf{Pros:}
 \begin{itemize}
\item Simplicity: Distance Vector Routing is straightforward to implement and understand, making it suitable for smaller or less complex networks.

\item Decentralization: The algorithm operates in a completely decentralized manner, where each router makes independent decisions based on the information from its neighbors.

\item Autonomy: It allows routers to dynamically adapt to changes in the network topology, such as link failures or the addition of new routers, without requiring a central authority.

 \end{itemize}
 \textbf{Cons:}
 \begin{itemize}
\item Slow Convergence: Distance Vector Routing can be slow to converge to a stable set of routing tables after a change in network topology. This can lead to temporary routing loops and count-to-infinity problems.

\item Scalability Issues: The periodic updates and the need to broadcast the entire routing table to neighbors can lead to scalability issues in larger networks, as it consumes bandwidth and processing power.

\item Security Vulnerabilities: Without proper security measures, malicious routers can introduce incorrect routing information into the network, leading to routing loops or black holes.
 \end{itemize}
 \begin{figure}[!t]
	\centering
	\includegraphics[width=2.5in]{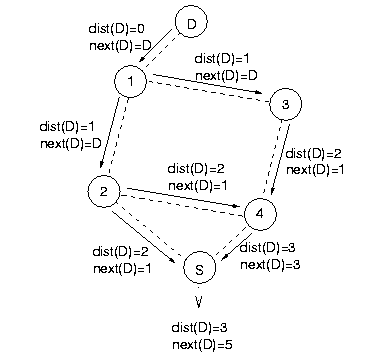}
	\caption{Example of Distance Vector Routing.\cite{vector}}
	\label{Example-of-Distance-Vector-Routing}
\end{figure}

 \textbf{Application:}
 \begin{itemize}
\item Small to Medium-Sized Networks: Due to its simplicity and ease of configuration, Distance Vector Routing is well-suited for small to medium-sized networks.

\item RIP Protocol: The Routing Information Protocol (RIP) is one of the most well-known implementations of Distance Vector Routing, used in many small to medium-sized networks.
 \end{itemize}
 Distance Vector Routing provides a simple and effective method for routing in packet-switched networks, with the benefit of decentralized decision-making and adaptability to network changes. However, its limitations in terms of convergence speed, scalability, and security need to be considered when choosing it for larger or more dynamic network environments. Despite these challenges, Distance Vector Routing remains foundational to understanding routing protocols and is an integral part of network design and operation in specific contexts.
\subsection{Link State Routing}
Link State Routing is an advanced network routing protocol that overcomes some of the limitations inherent in distance vector routing protocols. It operates by having each router learn the entire network topology, allowing for a more comprehensive and dynamic understanding of the network. Each router independently calculates the shortest path to every possible destination using its knowledge of the network topology. The algorithm most commonly associated with link state routing is Dijkstra's shortest path algorithm.\\
How Link State Routing Works:
 \begin{itemize}
\item Topology Discovery: Each router in the network discovers its immediate neighbors and learns the cost (e.g., bandwidth, delay, or other metrics) to reach them, typically through the exchange of hello packets.

\item Link State Advertisement (LSA): Every router creates a packet of information called a Link State Advertisement that contains its connectivity and the cost to each of its neighbors. These LSAs are then flooded to all routers in the network, ensuring that every router receives an identical copy of the topology information.

\item Topology Database: Each router uses the LSAs to build a complete map of the network's topology, stored in a database. This database is identical on all routers, providing a comprehensive view of the network.

\item Route Calculation: With the complete network topology at its disposal, each router independently calculates the shortest path to every destination using Dijkstra's algorithm. The outcome of this calculation is a shortest-path tree with the router itself as the root.
 \end{itemize}
 \begin{figure}[!t]
	\centering
	\includegraphics[width=2.5in]{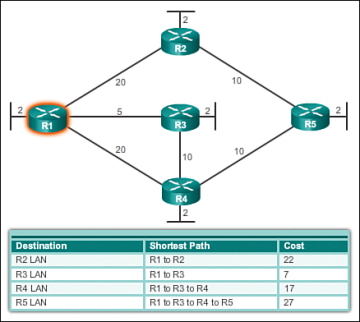}
	\caption{Link State Routing.\cite{linkstate}}
	\label{linkstate}
\end{figure}

 \textbf{Pros:}
\begin{itemize}
\item Rapid Convergence: Link state routing protocols converge more quickly than distance vector protocols because they have a complete view of the network. Changes are propagated rapidly, and each router independently recalculates routes, minimizing the time to convergence.

\item Scalability: The use of LSAs and the flooding mechanism allows link state protocols to scale effectively to larger networks compared to distance vector protocols.

\item Avoidance of Routing Loops: The global knowledge of the network prevents the formation of routing loops, a common problem in distance vector routing protocols.

\item Accurate and Efficient Routing: Having a complete view of the network allows routers to make informed decisions about the shortest paths, leading to more efficient use of network resources.

\end{itemize}
\textbf{Cons:}
\begin{itemize}
\item Resource Intensity: Maintaining a complete topology map and running Dijkstra's algorithm requires more memory and processing power than distance vector routing protocols.

\item Initial Traffic Overhead: The initial flooding of LSAs to establish the topology database can create significant traffic, although this is mitigated in steady-state operations.

\item Complexity: The mechanisms of LSA flooding, maintaining the topology database, and calculating routes add complexity to the router's operation and configuration.
 \end{itemize}
\textbf{Application:}
\begin{itemize}
\item OSPF (Open Shortest Path First): One of the most widely used link state routing protocols in IP networks. OSPF is designed for scalability and efficiency in diverse network topologies.

\item IS-IS (Intermediate System to Intermediate System): Another link state protocol used in large enterprise and service provider networks. It functions similarly to OSPF but can operate in both IP and OSI environments.

 \end{itemize}
Link State Routing provides a robust framework for routing in complex and diverse network environments, offering advantages in terms of convergence speed, scalability, and routing efficiency. Despite its resource requirements and operational complexity, the benefits of link state routing make it a preferred choice for large, dynamic networks seeking reliable and efficient routing solutions.
\subsection{Multipath Routing}

Multipath Routing is a network routing technique that finds multiple feasible paths between a source and a destination in a network. Unlike traditional routing methods that use a single best path for packet delivery, multipath routing leverages the redundancy of network paths to improve reliability, bandwidth, and load balancing. This approach can significantly enhance the overall performance and fault tolerance of network communications.

How Multipath Routing Works:
 \begin{itemize}
\item Path Discovery: Multipath routing protocols begin by discovering multiple paths between the source and destination. This can be achieved through various mechanisms, including extending traditional routing protocols or using specialized multipath discovery processes.

\item Path Selection: Among the discovered paths, the protocol selects a subset for use in routing. Selection criteria may include path length, bandwidth, latency, or other quality-of-service (QoS) metrics. The objective is to optimize the network's performance while avoiding congestion and ensuring reliability.

\item Traffic Distribution: Traffic is distributed across the selected paths based on predefined rules or dynamically in response to network conditions. Distribution strategies can range from simple round-robin to more complex algorithms that consider path characteristics and network load.

\item Path Maintenance: Multipath routing protocols monitor the status of active paths and can dynamically adjust the set of paths in use. If a path becomes unavailable or suboptimal due to network changes, the protocol can reroute traffic to maintain performance and reliability.
\end{itemize}
 \textbf{Pros:}
\begin{itemize}
\item Increased Bandwidth: By utilizing multiple paths simultaneously, multipath routing can aggregate bandwidth, offering higher data throughput.

\item Improved Reliability and Fault Tolerance: The availability of alternative paths enhances network reliability, as failures in one path can be mitigated by rerouting traffic through others.

\item Enhanced Load Balancing: Distributing traffic across multiple paths helps in balancing load more evenly across the network, preventing congestion on any single path.

\item Reduced Latency: In some cases, multipath routing can reduce latency by selecting the fastest path available for critical traffic.

\end{itemize}
\textbf{Cons:}
\begin{itemize}
\item Complexity: Implementing and managing multipath routing adds complexity to network design and operation, requiring sophisticated algorithms for path discovery, selection, and maintenance.

\item Overhead: The process of maintaining multiple paths and distributing traffic can introduce additional control traffic and processing overhead.

\item Interference and Out-of-Order Delivery: When packets for the same transmission are sent over paths with varying latencies, it can lead to packet reordering, requiring additional mechanisms to reorder packets at the destination.
 \end{itemize}
\textbf{Application:}
\begin{itemize}
\item Data Center Networks: Multipath routing is extensively used in data centers to optimize the use of available bandwidth and enhance fault tolerance.

\item Internet Traffic: Protocols like Multiprotocol Label Switching (MPLS) can implement multipath routing to improve the efficiency and reliability of internet traffic routing.

\item Wireless Networks: In wireless mesh networks and mobile ad hoc networks (MANETs), multipath routing can improve resilience and performance in the face of dynamic network conditions.

 \end{itemize}
Multipath Routing offers a powerful approach to enhancing network performance, reliability, and efficiency. By intelligently leveraging multiple paths through the network, it provides significant benefits over traditional single-path routing techniques. However, the increased complexity and overhead associated with managing multiple paths are important considerations when implementing multipath routing in a network.

\subsection{Energy-Efficient Routing}
Energy-Efficient Routing is a critical concept in the design and operation of wireless sensor networks (WSNs), ad hoc networks, and other types of networks where energy conservation is a priority. The primary goal of this routing strategy is to minimize energy consumption across the network to prolong the lifetime of battery-powered nodes, ensuring network sustainability and reliability over time. This is particularly important in environments where replacing or recharging batteries is impractical or impossible.

Principles of Energy-Efficient Routing:
 \begin{itemize}
\item Energy Awareness: Routing decisions are made based on the energy consumption of different paths and the residual energy of nodes, prioritizing routes that minimize overall energy usage and balance the energy expenditure among nodes to avoid early depletion of any single node.

\item Low Power Operations: The protocol may implement mechanisms to reduce power consumption, such as reducing transmission power, using energy-efficient data aggregation techniques, and putting nodes into sleep mode when they are inactive.

\item Multi-Hop Routing: Instead of direct communication between distant nodes and a base station, which can be energy-intensive, data packets are forwarded through multiple intermediate nodes using shorter, energy-efficient hops.
\end{itemize}
Strategies for Energy-Efficient Routing
 \begin{itemize}
\item Minimum Energy Routing: Selects the path that consumes the least amount of energy, regardless of the number of hops, to transfer data from the source to the destination.

\item Energy Balancing: Focuses on distributing the energy consumption evenly across the network to prevent the rapid depletion of individual nodes, which could create network gaps.

\item Data Aggregation: Involves combining data from multiple nodes to reduce the number of transmissions, thereby saving energy. This is particularly effective in networks where data from different sensors can be aggregated without loss of information.

\end{itemize}

\begin{figure}[!t]
	\centering
	\includegraphics[width=2.5in]{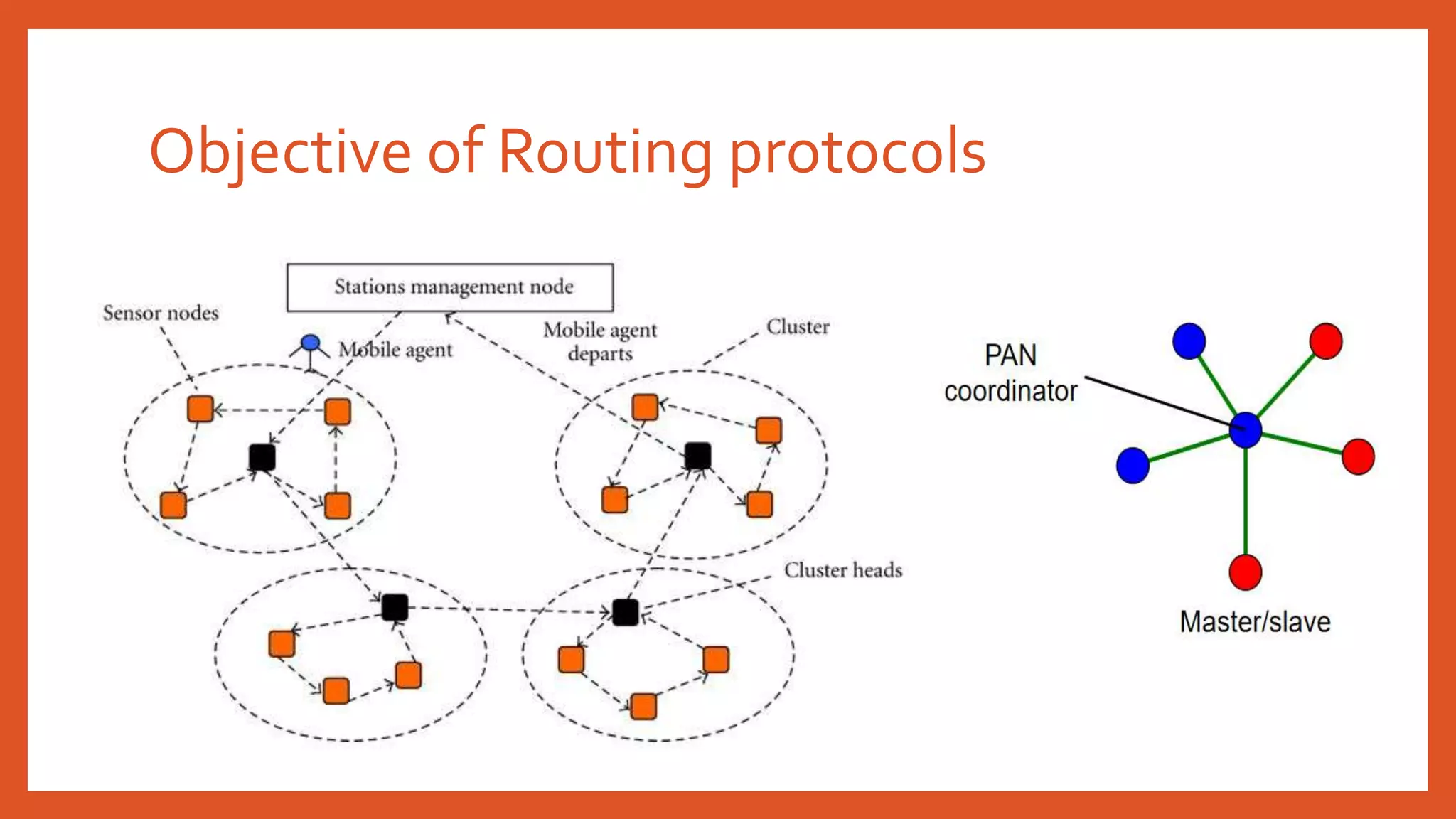}
	\caption{Energy Efficient Routing\cite{energy}}
	\label{wsnrouting-protocols-energy-efficient-routing-3-2048}
\end{figure}
\textbf{Pros:}
\begin{itemize}
\item Extended Network Lifetime: By minimizing energy consumption, energy-efficient routing protocols significantly extend the operational lifespan of battery-powered networks.
\item Increased Reliability: Evenly distributing energy usage helps prevent node failures, enhancing the overall reliability of the network.
Cost Efficiency: Reducing the frequency of battery replacements lowers the maintenance cost of deploying wireless sensor networks in remote areas.
\end{itemize}
\textbf{Cons:}
\begin{itemize}
\item Complexity: Energy-efficient routing algorithms can be more complex to implement, requiring additional computations to assess energy metrics and make routing decisions.
\item Performance Trade-offs: Focusing on energy efficiency might result in increased latency or reduced data throughput due to the preference for multi-hop routing over more direct paths.
\item Dynamic Network Conditions: The effectiveness of energy-efficient routing can be influenced by changes in network topology, such as node mobility or failure, requiring adaptive algorithms to maintain efficiency.
 \end{itemize}
\textbf{Application:}
\begin{itemize}
\item Wireless Sensor Networks (WSNs): Used in environmental monitoring, military surveillance, and smart agriculture, where sensors are often deployed in inaccessible locations.
\item Internet of Things (IoT): In IoT applications, extending battery life is crucial for devices that are expected to operate for long periods without human intervention.
\item Ad Hoc Networks: In scenarios like disaster recovery or military operations, where network infrastructure might be unavailable or impractical, energy efficiency ensures longer operational periods.
 \end{itemize}
Energy-Efficient Routing is essential for optimizing the longevity and reliability of networks where energy resources are limited. Through careful routing decisions and techniques aimed at minimizing and balancing energy consumption, these protocols help to ensure that networked systems, especially those deployed in challenging or remote environments, can continue to operate effectively over extended periods. Despite the challenges and trade-offs involved, the benefits of energy-efficient routing in terms of cost savings, network sustainability, and device longevity are significant, making it a key consideration in the design of modern wireless networks.

\section{Some modern Routing algorithm}
\subsection{EESRA\cite{8765561}}
\subsubsection{Introduction}
This paper introduces an energy-efficient clustering and hierarchical routing algorithm named EESRA, aimed at extending the lifespan of wireless sensor networks (WSNs) with increasing network size. This algorithm is an enhancement over the "low-energy adaptive clustering hierarchy" (LEACH) protocol, addressing its scalability issues by adopting a three-layer hierarchy to minimize cluster heads' load and randomize their selection. Moreover, EESRA employs multi-hop transmissions for intra-cluster communications to implement a hybrid WSN MAC protocol. The paper's simulation results demonstrate that EESRA outperforms other WSN routing protocols in terms of load balancing and energy efficiency on large-scale WSNs.

The key contributions of the paper include:
\begin{itemize}
  \item An innovative energy-efficient clustering and hierarchical routing algorithm that outperforms existing WSN routing protocols in terms of scalability and energy efficiency.
  \item The introduction of a hybrid WSN MAC protocol that incorporates both sleep and collision avoidance mechanisms alongside TDMA slots for efficient data forwarding.
  \item A comprehensive simulation analysis demonstrating the algorithm's effectiveness in extending the network lifespan and achieving better load balancing across large-scale WSNs.
\end{itemize}
\begin{figure*}[!t]
	\centering
	\subfloat[]{\includegraphics[width=2.5in]{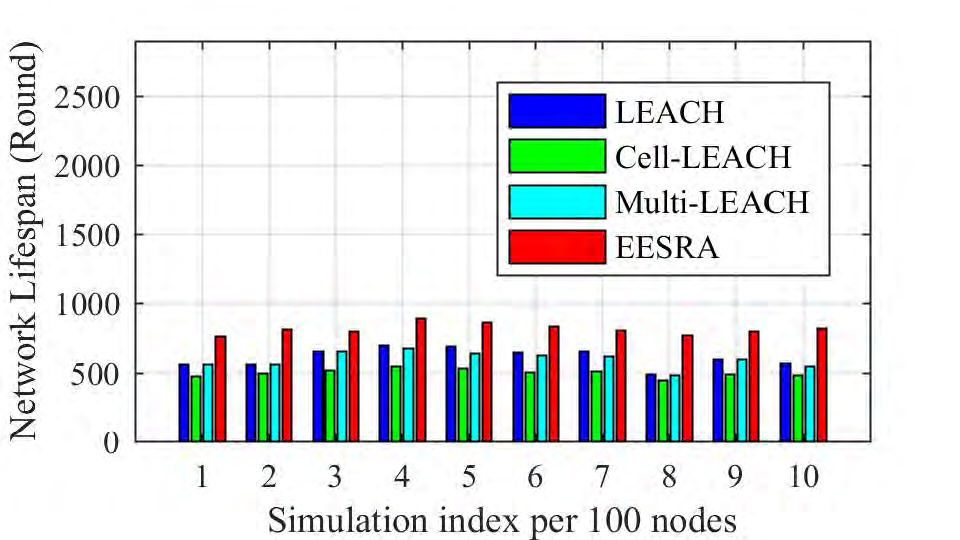}%
	\label{fig_first_case}}
	\hfil
	\subfloat[]{\includegraphics[width=2.5in]{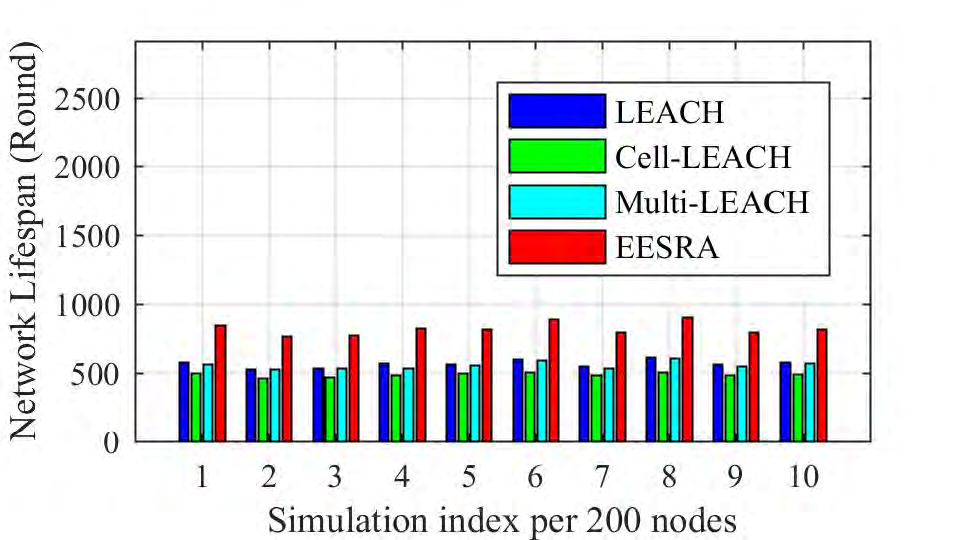}%
	\label{fig_second_case}}
	\hfil
	\subfloat[]{\includegraphics[width=2.5in]{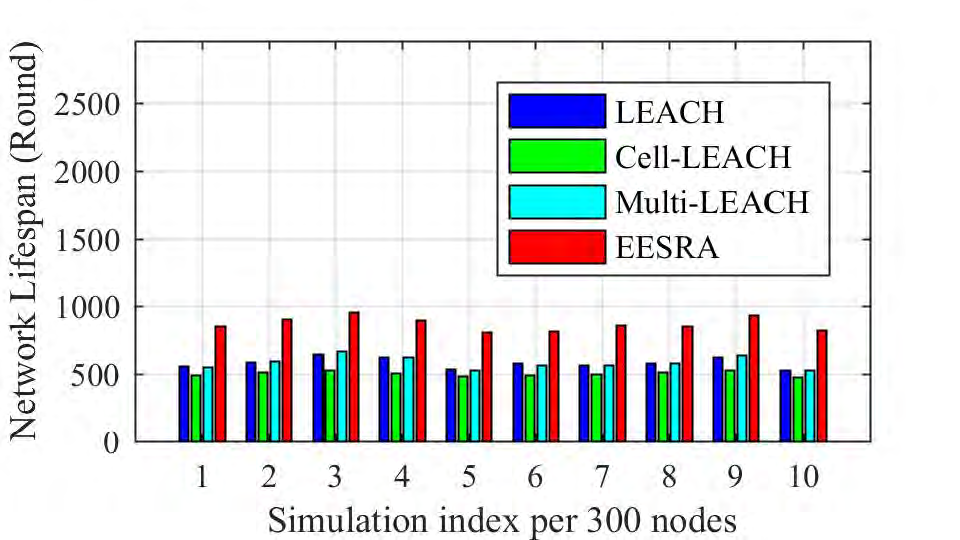}%
	\label{fig_third_case}}
	\hfil
	\subfloat[]{\includegraphics[width=2.5in]{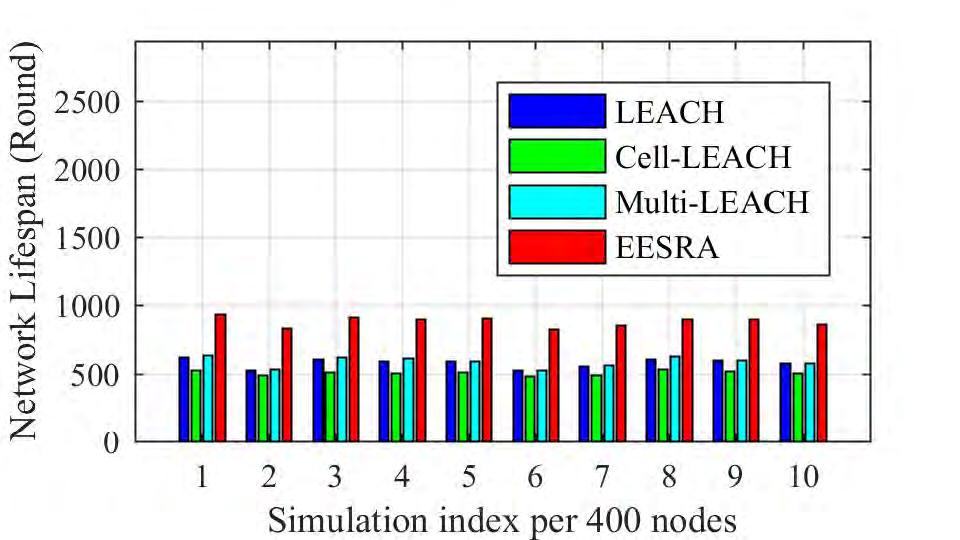}%
	\label{fig_fourth_case}}
	\caption{Packets delivery ratio per:  (a) 100 nodes, (b) 200 nodes, (c) 300 nodes, (d) 400 nodes\cite{8765561}}
	\label{fig_sim}
	\end{figure*}
\subsubsection{Optimization}
The EESRA (Energy Efficient Scalable Routing Algorithm) optimizes traditional energy-efficient routing in several key ways, focusing on addressing the scalability and energy consumption issues inherent in the LEACH protocol and other WSN (Wireless Sensor Network) routing protocols:
\begin{itemize}
	\item Three-layer Hierarchy: EESRA introduces a three-layer hierarchical structure to reduce the load on cluster heads (CHs) by incorporating an additional layer of cluster congregations (CGs) between the CHs and the cluster members (CMs). This structure helps in distributing the energy consumption more evenly across the network, thus optimizing energy usage and extending the network lifespan.

	\item Hybrid MAC Protocol: The algorithm employs a hybrid Medium Access Control (MAC) protocol that combines sleep and collision avoidance mechanisms for efficient data sensing and transmission. This approach allows for more efficient use of energy, as nodes can enter a low-power state when not actively transmitting data.

	\item Multi-hop Intra-cluster Communication: Unlike traditional protocols that may rely solely on single-hop communication, EESRA uses multi-hop transmissions within clusters. This method reduces the energy consumed in data transmission by minimizing the distance over which individual transmissions need to occur.

	\item Randomized Cluster Head Selection: To further balance the energy consumption across the network, EESRA implements a randomized selection of cluster heads. This ensures that no single set of nodes bears the brunt of the energy consumption, leading to a more uniform depletion of resources across the network.

Load Balancing: By adopting a three-layer hierarchy and utilizing multi-hop communication, EESRA effectively balances the load among nodes. This prevents certain nodes from depleting their energy resources too quickly, thereby optimizing the overall energy usage within the network.
\end{itemize}
\subsubsection{Performance Evaluation}
The simulation results Fig.\ref{fig_sim} presented in the paper demonstrate that EESRA outperforms traditional LEACH and its variants in terms of energy efficiency, load balancing, and scalability, especially in large-scale WSNs. By addressing the critical challenges of energy consumption and network scalability, EESRA optimizes traditional energy-efficient routing protocols, offering a significant improvement in extending the network lifespan while maintaining high network performance.

\begin{table}[h]
	\centering
	\caption{Comparison of LEACH, Cell-LEACH, and EESRA Protocols\cite{8765561}}
	\label{tab:comparison}
	\begin{tabular}{|l|l|c|c|c|c|}
	\hline
	\textbf{Nodes} & \textbf{Protocol} & \textbf{FDN} & \textbf{FDN Ratio} & \textbf{AND} & \textbf{AND Ratio} \\ \hline
	100 & LEACH&42&1&613&1\\ & Cell-LEACH&129&3.14&501&0.82\\ &Multi-LEACH&107&2.6&597&0.97\\ &EESRA&363&8.85&819&1.33                              \\ \hline
	200 & LEACH&41&1&570&1\\ & Cell-LEACH&122&2.97&490&0.86\\ &Multi-LEACH&123&3&559&0.97\\ &EESRA&330&8.05&826&1.45                    \\ \hline
	300 & LEACH&41&1&585&1\\ & Cell-LEACH&129&2.97&505&0.86\\ &Multi-LEACH&122&2.97&597&1\\ &EESRA&322&7.85&872&1.49                     \\ \hline
	400 & LEACH&42&1&581&1\\ & Cell-LEACH&129&3.14&507&0.87\\ &Multi-LEACH&121&2.95&597&1.01\\ &EESRA&312&7.61&883&1.52                       \\ \hline
	\end{tabular}
\end{table}

\subsection{GAPSO‐SVM\cite{NorouziShad2022}}
\subsubsection{Introduction}
GAPSO-SVM is proposed as an innovative clustering routing protocol, specifically designed for the IoT perception layer, with an emphasis on energy-aware localization and routing. It integrates a Support Vector Machine (SVM) for precise location estimation and a Genetic Algorithm-Particle Swarm Optimization (GAPSO) for efficient clustering. The primary contributions of this approach include:

\begin{enumerate}
    \item A hybrid IDSS-based clustering routing protocol that significantly enhances energy efficiency and network longevity over previous methodologies.
    \item An SVM-based localization algorithm that enables accurate location estimation without the need for GPS, addressing a common limitation in geographic protocols.
    \item A hybrid GAPSO algorithm that optimizes clustering with superior convergence rate and efficiency compared to similar endeavors.
\end{enumerate}

The GAPSO-SVM algorithm's framework involves specifying sensor nodes and beacons, alongside calculating the energy required for transmitting and receiving data, optimizing the network's energy consumption for effective data communication from cluster heads (CHs) to the sink.

Through simulation, GAPSO-SVM demonstrated substantial improvements in network lifetime and energy efficiency, utilizing SVM for precise localization without GPS and leveraging GAPSO for efficient clustering. The results indicated marked advancements in convergence rates, network longevity, and energy savings, significantly outperforming the metrics of the existing EEWC algorithm.
\subsubsection{Optimization Process}

The GAPSO-SVM algorithm employs a hybrid optimization strategy that combines the strengths of Genetic Algorithms (GA) and Particle Swarm Optimization (PSO) to enhance the clustering and routing processes in IoT networks. The optimization process is detailed as follows:

\begin{enumerate}
    \item \textbf{Initialization:} The set of sensor nodes within the network is defined, alongside the specification of beacon and non-beacon nodes. This step forms the basis for the clustering and routing mechanism.
    
    \item \textbf{Energy Calculation:} For each node, the energy required to transmit (ET) and receive (ER) an L-bit message over a distance \(d\) is calculated using:
    \begin{equation}
    ET = \begin{cases} 
    L \cdot E_{elec} + L \cdot \epsilon_{fs} \cdot d^2 & \text{if } d < d_0 \\
    L \cdot E_{elec} + L \cdot \epsilon_{mp} \cdot d^4 & \text{otherwise}
    \end{cases}
    \end{equation}
    \begin{equation}
    ER = L \cdot E_{elec}
    \end{equation}
    where \(E_{elec}\) is the energy consumed per bit by the transmitter or receiver circuit, and \(\epsilon_{fs}\), \(\epsilon_{mp}\) are the amplifier energy consumption parameters for free space and multipath models respectively.
    
    \item \textbf{Hybrid GAPSO Mechanism:} The optimization leverages the fast convergence rate of PSO and the robust search capabilities of GA. The population of solutions (sensor nodes' clustering configurations) is evolved using both GA and PSO principles:
    \begin{itemize}
        \item Part of the population is processed using GA operations (selection, crossover, and mutation) to explore the search space.
        \item The remaining part is updated using PSO rules, guiding the particles (solutions) toward the best-found positions.
    \end{itemize}
    
    \item \textbf{Hybridization Coefficient (HC):} A key parameter in GAPSO, HC determines the proportion of the population to be processed by GA in each iteration. An optimal HC value ensures a balanced exploration and exploitation, enhancing the algorithm's efficiency and convergence.
    
    \item \textbf{Evaluation and Iteration:} The fitness of each solution is evaluated based on criteria such as energy efficiency, network lifetime, and connectivity. The population is then updated iteratively, combining GA and PSO updates to find an optimal clustering configuration.
\end{enumerate}
\begin{figure}
    \centering
    \subfloat[Comparison of GA and GAPSO cost function for each iteration\cite{NorouziShad2022}]{\includegraphics[width=2.5in]{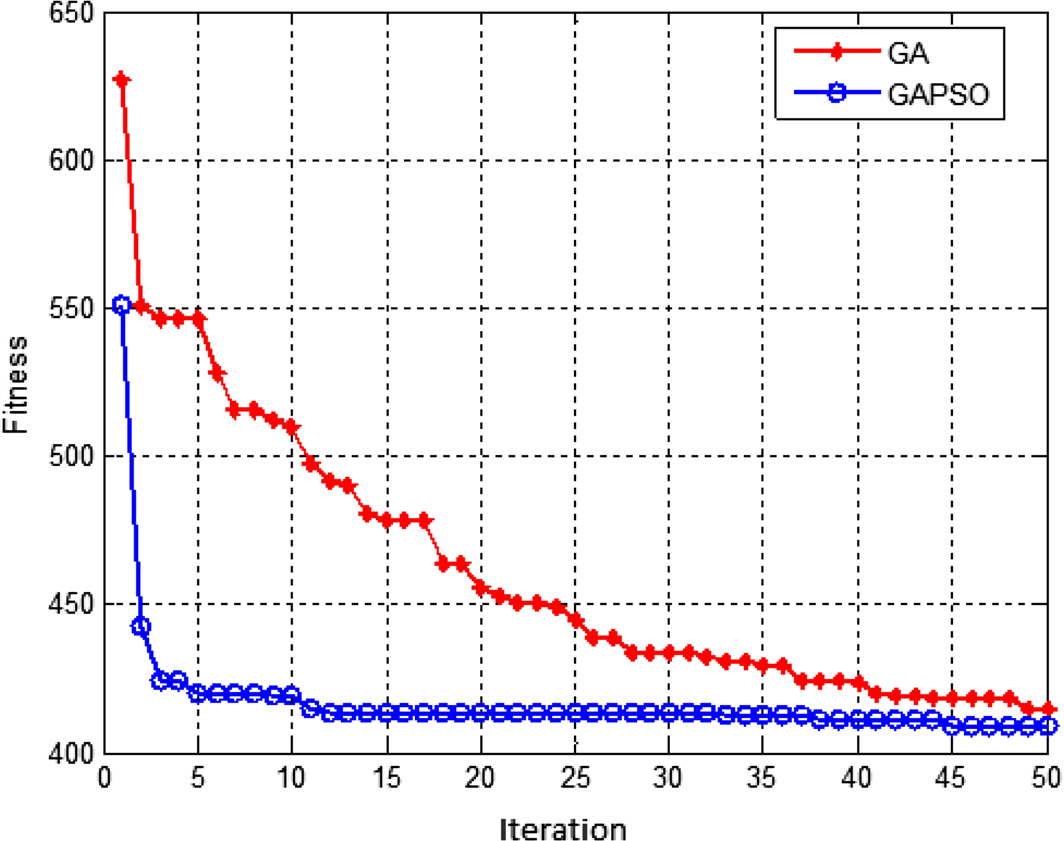}\label{fig_convergence}}
    \hfill 
    \subfloat[Network lifetime for 10\% beacon nodes\cite{NorouziShad2022}]{\includegraphics[width=2.5in]{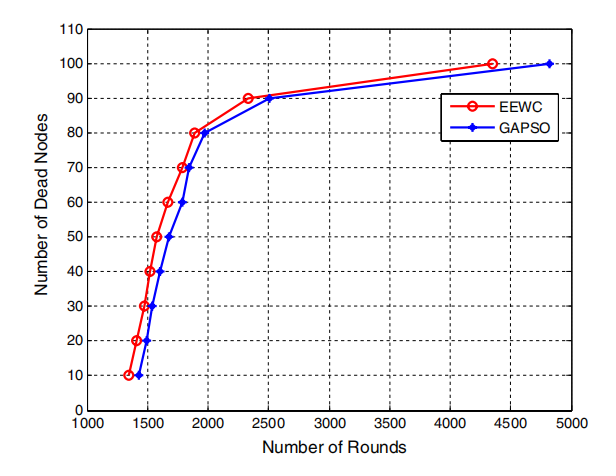}\label{fig_lifetime}}
    \hfill 
    \subfloat[Comparison of residual energy in GAPSO and EEWC algorithms for 10\% beacon node\cite{NorouziShad2022}]{\includegraphics[width=2.5in]{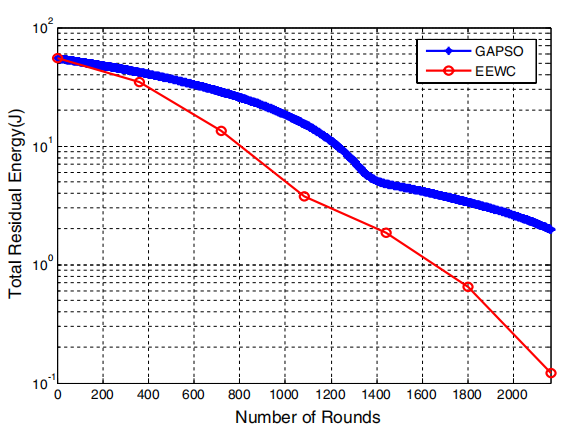}\label{fig_energy}}
    \caption{Various performance evaluation. (a) Cost function convergence, (b) Network lifetime, and (c) Residual energy comparison.}
\end{figure}
\subsubsection{Performance Evaluation}
The performance of GAPSO-SVM was rigorously tested against the EEWC algorithm, showcasing significant improvements in network lifetime and energy efficiency. Key findings include:
\begin{itemize}
    \item \textbf{Improved Convergence Rate:} GAPSO-SVM demonstrated an 80\% improvement in convergence rate over EEWC(Fig.\ref{fig_convergence}), leading to faster optimization of CHs.
    \item \textbf{Extended Network Lifetime:} The network lifetime under GAPSO-SVM extended by 11\% (Fig.\ref{fig_lifetime}), attributed to optimized clustering and routing strategies.
    \item \textbf{Energy Efficiency:} GAPSO-SVM reduced the energy consumption significantly(Fig.\ref{fig_energy}), ensuring a sustainable IoT network operation.
\end{itemize}
\subsubsection{Conclusion}
The simulation results indicate that GAPSO-SVM outperforms the existing EEWC algorithm in terms of convergence rate, network longevity, and energy efficiency, making it a promising solution for energy-aware localization and routing in IoT networks.
\bibliographystyle{IEEEtran}
\bibliography{my_essay}

\begin{thebibliography}{1}
\providecommand{\url}[1]{#1}
\csname url@samestyle\endcsname
\providecommand{\newblock}{\relax}
\providecommand{\bibinfo}[2]{#2}
\providecommand{\BIBentrySTDinterwordspacing}{\spaceskip=0pt\relax}
\providecommand{\BIBentryALTinterwordstretchfactor}{4}
\providecommand{\BIBentryALTinterwordspacing}{\spaceskip=\fontdimen2\font plus
\BIBentryALTinterwordstretchfactor\fontdimen3\font minus
  \fontdimen4\font\relax}
\providecommand{\BIBforeignlanguage}[2]{{%
\expandafter\ifx\csname l@#1\endcsname\relax
\typeout{** WARNING: IEEEtran.bst: No hyphenation pattern has been}%
\typeout{** loaded for the language `#1'. Using the pattern for}%
\typeout{** the default language instead.}%
\else
\language=\csname l@#1\endcsname
\fi
#2}}
\providecommand{\BIBdecl}{\relax}
\BIBdecl

\bibitem{shortest}
wikipedia, ``{Shortest path problem},''
  \url{https://en.wikipedia.org/wiki/Shortest_path_problem}, 2024, [Online;
  accessed 22-Feb-2024].

\bibitem{dimension}
M.~Behrouzian~Nejad, A.~Mehranzadeh, and M.~Hoodgar, ``Performance of input and
  output selection techniques on routing efficiency in network-on-chip,''
  \emph{International Journal of Computer Science and Information Security,},
  vol.~9, 09 2011.

\bibitem{9796682}
S.~Supittayapornpong, P.~Namyar, M.~Zhang, M.~Yu, and R.~Govindan, ``Optimal
  oblivious routing for structured networks,'' in \emph{IEEE INFOCOM 2022 -
  IEEE Conference on Computer Communications}, 2022, pp. 1988--1997.

\bibitem{ABDALLAH2016264}
\BIBentryALTinterwordspacing
A.~E. Abdallah, ``Smart partial flooding routing algorithms for 3d ad hoc
  networks,'' \emph{Procedia Computer Science}, vol.~94, pp. 264--271, 2016,
  the 11th International Conference on Future Networks and Communications (FNC
  2016) / The 13th International Conference on Mobile Systems and Pervasive
  Computing (MobiSPC 2016) / Affiliated Workshops. [Online]. Available:
  \url{https://www.sciencedirect.com/science/article/pii/S1877050916317872}
\BIBentrySTDinterwordspacing

\bibitem{vector}
A.~Gupta, A.~Smith, and A.~Sun, ``Design and implementation of fisheye routing
  protocol for mobile ad hoc networks,'' \emph{none}, 09 2001.

\bibitem{linkstate}
cisco, ``{Link State Routing Protocol},''
  \url{https://www.ciscopress.com/articles/article.asp?p=2180210&seqNum=11},
  2014, [Online; accessed 22-Feb-2024].

\bibitem{energy}
ArunChokkalingam, ``{Energy Efficient Routing},''
  \url{https://www.slideshare.net/ArunChokkalingam/wsnrouting-protocols-energy-efficient-routing},
  2020, [Online; accessed 22-Feb-2024].

\bibitem{8765561}
E.~F. Ahmed~Elsmany, M.~A. Omar, T.-C. Wan, and A.~A. Altahir, ``Eesra: Energy
  efficient scalable routing algorithm for wireless sensor networks,''
  \emph{IEEE Access}, vol.~7, pp. 96\,974--96\,983, 2019.

\bibitem{NorouziShad2022}
\BIBentryALTinterwordspacing
M.~N. Shad, M.~Maadani, and M.~N. Moghadam, ``Gapso-svm: An idss-based
  energy-aware clustering routing algorithm for iot perception layer,''
  \emph{Wireless Personal Communications}, vol. 126, no.~3, pp. 2249--2268,
  2022. [Online]. Available: \url{https://doi.org/10.1007/s11277-021-09051-5}
\BIBentrySTDinterwordspacing

\end{thebibliography}

\end{document}